\documentclass[aps,prl,twocolumn]{revtex4}
\usepackage{bm}
\usepackage{epsfig} 
\usepackage{amsfonts}
\usepackage{amsmath}

\allowdisplaybreaks[1] 

\newcommand*{\be}{\begin{equation}}
\newcommand*{\ee}{\end{equation}}
\newcommand*{\bea}{\begin{eqnarray}}
\newcommand*{\eea}{\end{eqnarray}}

\begin{document}


\title{Explaining the Magnetic Moment Reduction of   Fullerene encapsulated Gadolinium through a Theoretical Model  }

\author{Alessandro Mirone}
\affiliation{\dag European Synchrotron Radiation Facility, BP 220, F-38043 Grenoble Cedex, France}

\date{\today}

\begin{abstract}
We propose a Theoretical model accounting for the recently observed reduced magnetic moment of Gadolinium in fullerenes. While this reduction has been observed also for other  
trivalent rare-hearth atoms (Dy$^{3+}$, Er$^{3+}$, Ho$^{3+}$) in
fullerenes and can be ascribed to crystal field effects,  the explanation of this phenomena for Gd$^{3+}$ is not straightforward  due to the  sphericity of its ground state ( S$=7/2$ , L$=0$ ).
In our model the  momentum lowering is the result of a subtle interplay between hybridisation and spin-orbit interaction.
\end{abstract}


\maketitle

\section{INTRODUCTION}

Endohedral metallofullerenes M@$C_{82}$ are novel materials  that have attracted a wide interest in physics, chemistry but also in material or biological sciences for the large variety of promising applications of their peculiar properties\cite{shi1,bet1,she1}. 
In endohedral metallofullerenes, a positively charged 
core metal is off-center in a negatively charged strong carbon cage, resulting in strong metal-cage interaction and 
intrafullerene charge transfer from the metal to the cage.\cite{shi1,poi1,gie1,iid1}
 
The magnetism of these systems is mainly due to the spin of the entrapped
metals. In a series of average magnetization mesurements  a paramagnetic
behaviour has been observed \cite{fun1,nut1,dun1, hua1}, with negative Weiss
temperatures. The negative Weiss temperature    indicates
the presence of a weak  antiferromagnetic interaction betwen the cage
and the metal,  and between neighbooring cages,  but
for heavy rare-earths (RE) endofullerenes\cite{hua1} ferromagnetic
coupling  has been mentioned in the  sub-20K range. 
In the case of  heavy RE these experiments gave a number of magnetons per encaged ion that is
smaller than for the free ion. This result has been
phenomenogically ascribed to the cage  crystal field interaction for high L
ions and, for the $L=0$ Gd case, to the antiferommagnetic interaction
between the ion and the cage.  

A recent work\cite{denadai} has used x-ray magnetic circular dichroism and (XMCD) x-ray absorption spectroscopy (XAS)
to  characterize  local magnetic properties of heavy RE metallo centers, using the 
$M_4$ and $M_5$ resonances ($3d \rightarrow 4f$ transitions). The absorption spectra of this work were very well fitted
assuming trivalent ions ($4f^n$ electronic structure with $n=7$ for Gd, $n=9$ for Dy...), while  XMCD
confirmed that there is  a strong reduction of the measured ion  magnetisation   compared to the free ion case.

For $L\ne 0$ ions the reduction was reproduced by a model Hamiltonian were a weak crystal field 
prevents the ion total angular moment $J$ to align completely along the magnetic field.
The case $L=0$ of trivalent Gadolinium was more difficult and hybridisation model gave not
a satisfactory explanation.

Infact, althought hybridisation gives antiferromagnetic coupling with the cage and accounts ( in the Gd case) for a 14\% 
reduction of the average moment ( Gd + cages), it  cannot  explain the
reduction of the  moment localized on  Gd ion (see next section).

What we will show in the present paper is that a combined action of hybridisation and spin-orbit interaction 
can have a dramatic effect on the observed magnetic moment. 

This effect is not trivial and is significant only in a 
restraint parameters region that was not discovered in the previous numerical study\cite{denadai}.
In the present paper we will give  a complete  analytical discussion of this effect.
In next section  we will introduce a simple model with an anisotropic hybridisation  where the Gd ground
state is basically $4f^7$ with a small $4f^8$ component due to a backdonation from the cage.
we will then consider  spin-orbit interactions at the first order and show the dramatic changes in magnetisation.
 Analytical formula will  be compared to exact numerical solutions. Finally  we will discuss the possible
experimental manifestations of the studied phenomena.

\section{Hybridisation Model and spin-orbit}

Back donation concerns a cage unpaired electron. 
As the Re ion is off center we consider that the  
hopping is non isotropic and choose to restrain the  transfer of  the  cage backdonated electron  to the $4f$ orbital that is closest to the cage, the $m_z=0$ one, where the z-axis is parallel to the off center displacement.

The Gd ground state has $S=7/2$, $L=0$. Adding one more electron to this state, one can acces 
only the quantum numbers $S=3$, $L=3$ 
of the $4f^8$ configuration\cite{rachaIII}.

The energy difference between the  $4f^8$ level and $4f^7$ ground
state will be named, without SO coupling, $\Delta E$. This is a positive
scalar quantity and must be  large in magnitude, compared to the hopping strenght  $t$, because the fractional backdonation
has been observed to be very small.

We consider an  effective interaction  term proportional to $t^2$. 
This interaction transforms the state ($S_z,s$) (where $S_z$ is the
spin z-component  of the $4f^7$ shell and s the cage unpaired electron
one) into the state  ($S_z+2s,-s$), and the other way round.

We can therefore restrict the  effective Hamiltonian to the
two-dimensional space spanned by ($S_z,s$) and ($S_z+2s,-s$ ). The elements of the effective interactions are :
\begin{eqnarray}  H(S_z,s \rightarrow S_z + s -s^{'}, s^{'}) =   \label{effective_simple}  \\
-t^2 \sum_{\eta}  \frac{  \langle  S_z + s -s^{'} | c_{0s^{'}} | \eta \rangle
 \langle \eta  | c^{+}_{0s} | S_z  \rangle 
      }{E_{\eta} -E_0}  \nonumber
\end{eqnarray}
where $E_0$ is the $4f^7$ ground state energy, $c^+_{ms}$ and $c_{ms}$ are creation/annihilation operators for 
a $4f$ electrons with $m_z=m$ and $s_z=s$. The sum runs over the states $\eta$  of the $4f^8$ configuration.

The case of Gd is quite simple because the only $4f^8$ state having a non-zero parentage coefficient with $|4f^7 S=7/2 L=0\rangle$
ground state is $ |4f^8 S^{'}=3, L^{'}=3 \rangle $. 

The results of operating on the ground state on $4f^7$
can be expressed in terms of the parentage coefficients and angular recoupling factors using well known formula\cite{sureau}, for our specific case:
\begin{eqnarray}
  c^+_{0s} |4f^7 S  S_z M_z \rangle = - \sqrt{8} G^{ S^{'}L^{'} }_{SL}
  \times \nonumber \\
(S S_z  1/2 s ; S^{'} (S_z+s) )  \times |4f^8 S L (S_z+s) M_z \rangle
\end{eqnarray}
Putting this formula into (\ref{effective_simple}) the Hamiltonian can be written as :
\begin{equation} H = -\frac{t^2}{\Delta E} \frac{8}{7} v_{S_z} \otimes v_{S_z} \label{Heffective_simple}
\end{equation}
where $v_{S_z}$ is the versor :
\begin{equation} v_{S_z} = ( (7/2+S_z)^{1/2}, -(7/2-S_z+1)^{1/2} )\frac{1}{\sqrt{8}} \label{simpleversor}\end{equation}
   
For positive $\Delta E$ 
   the ground state of H is $v_{S_z}$, it corresponds to antiferromagnetic alinement ( total angular moment
$J=3$), and 
has energy $-\frac{t^2}{\Delta E} \frac{8}{7}$.

The state perpendicular to $v_{S_z}$ has energy zero and corresponding total angular moment $J=4$.
There is no energy dependency on $S_z$ as it could have been expected on the basis of rotational invariance in 
the spin space, as long as spin-orbit interaction is not included. 
The local moment of Gd,  in the antiferromagnetic ground state, can be almost fully aligned along 
the magnetic field. On the basis of equation (\ref{simpleversor}) one should observe, at saturation
\begin{equation}
\langle S_z \rangle = (7/2 \times 7 + 5/2)/ 8 =3.375
\end{equation}
corresponding to a $3.6$\% reduction which is very far from the $20$\% observed one.
The antiferromagnetic metal-cage coupling cannot   explain the moment reduction observed with XAS, XMCD techniques.

Therefore we add the spin-orbit (SO) coupling to the picture   which breaks rotational invariance in the spin space.
At first order SO splits the energies of the $S^{'}=3$ $L^{'}=3$ state  according to the total $J^{'}$, and affects the denominator
involved in equation (\ref{effective_simple}).
As a result the equation  for H, that is given by equation \ref{Heffective_simple} for the zero-SO case, must be rewritten 
in this form :
\begin{equation} H_{SO}(S_z) = F(S_z ) v_{S_z} \otimes v_{S_z} \label{Heffective_simpleS}
\end{equation}
where 
\begin{equation}
   F(S_z ) = - t^2 \frac{8}{7} \sum_{J^{'}}  \frac{(2 J^{'} +1 )  (
\begin{array}{ccc} 
3 & 3 & J^{'}\\
S_z^{'} & 0 & -S_z^{'}
\end{array}
)^2}{\Delta E + E^{SO}_{J^{'}}} \label{firstordersimple}
\end{equation}

where $S_z^{'}=S_z-1/2$. From a formal point of view SO gives at first order also another contribution beside 
affecting the propagator denominator.  In fact the $4f^7$  ground state, considering SO interaction, is not
a pure $S=7/2$ , $L=0$ state, but it has also a small component of the
state $S=5/2$, $L=1$, whose amplitude is first order in SO strength. It is this perturbed ground state that should be considered in equation (\ref{effective_simple}).
However  for symmetry reasons, in the framework of our model where backdonation affects only the $m_z=0$ orbital, the contribution
at first order in SO coming from this  $S=5/2$, $L=1$ component is zero.

At the moment we will therefore restrain the discussion to the above equation (\ref{firstordersimple}),
that contains all the important physics of the studied phenomena.

The $S_z$ dependency in equation (\ref{firstordersimple}) is given by the $J^{'}$ dependecy of $ E^{SO}_{J^{'}}$.
If the denominator in equation (\ref{firstordersimple})  is constant, one can factor the term :
\begin{equation}   \sum_{J^{'}}  (2 J^{'} +1 )  (
\begin{array}{ccc} 
3 & 3 & J^{'}\\
S_z^{'} & 0 & -S_z^{'}
\end{array}
)^2 =1  \label{sumrule}\end{equation} and obtain again equation (\ref{Heffective_simple}). 
For the $4f^8 ,S^{'}=3,L^{'}=3 $ state the energy $E^{SO}_{J^{'}}$ is
\begin{equation}  E^{SO}_{J^{'}} = \zeta_{nl} \frac{24-J^{'}(J^{'}+1)}{12}  \label{esofirst} \end{equation}

where $\zeta_{nl}$ is the strength of SO interaction. This equation can be obtained in a simple way :  the state obtained putting  seven spin-up
electrons in the 4f shell, plus one spin-down electron in the $m_z=3$ state, has $S=3$, $L=3$, $J=6$.

It is very easy to calculate for this mono-determinantal state 
the expectation value of the $S.L$ scalar operator, it is $-1.5$ ( $m_z=3$ of the eighth electron times its spin ).
The expectation values for the other $J^{'}$ can be calculated observing that the expectation  value of 
a scalar product of two $L=1$ tensors must be proportional to the 6J
factor  for  the  angular moments sextet (3,3,$J^{'}$,1,1,0). By a quick glance at sixJ tables, equation (\ref{esofirst})
is readily obtained.

The energy correction $E^{SO}_{J^{'}}$, considering $\zeta_{nl}=0.1975eV$\cite{thole} has the
 negative value of about $-0.3$eV 
for $J^{'}=6$  and a positive value of about $0.4$eV for $J^{'}=0$. These values have to be compared to $\Delta E$.
Taking  $\Delta E$ of the order of $1eV$ the effect of $ E^{SO}_{J^{'}}$ is not negligeable
and the dependency on $S_z$ is given mainly by the term with the smaller denominator, the  $J^{'}=6$ term. 
The Wigner symbol 

\begin{equation}  ( \begin{array}{ccc} 
3 & 3 & 6\\
S_z^{'} & 0 & -S_z^{'}
\end{array})
 \end{equation} 
has the bigger value for  the smaller  $S_z^{'}$, as one can understand classically 
 considering that  to get $J^{'}=6$  one has to  align  $S^{'}$ and $L^{'}$.  

Therefore the ground state has the smallest $S^{'}_z$ and, unless the polarising magnetic field 
is sufficiently strong, the observed local magnetic moment will be
 zero.

The $J^{'}=6$ term is the only one to consider for $\Delta E$ approaching the value of $0.3$ eV, 
because its denominator in equation (\ref{firstordersimple}) tends to zero. But a small denominator means strong 
hybridisation, while  hybridisation is weak because the encaged Gd is an almost pure $4f^7$ 
configuration. 

One should therefore consider the  region where $\Delta E + E^{SO}_{J^{`}}$ is big compared 
to the hopping strenght.
For $\Delta E$ going to infinity the studied effect cancels out( see equation (\ref{sumrule})).
 
So we are going to study a region where the effect 
 results from a imperfect cancellation of the different terms involved in the sum of equation  (\ref{firstordersimple}).

As a cancellation is involved one has to be very precise evaluating each single term in the sum.
Therefore we devote a particular attention to the exact values of $E^{SO}_{J^{'}}$.

 These 
values could be obtained at second order using Racah formalism and summing contributions from
all the $4f^8$ states accessible operating with the spin-orbit interaction
on the    $4f^8,S=3,L=3$ state.
However, for the scope of this paper, which is to clarify the
consequences of equation (\ref{firstordersimple}), it is sufficient to plug in the sum the energies 
obtained by exact diagonalisation of $Gd^{2+}$ ion.

 We show in table (1) the comparison of 
$E^{SO}_{J^{'}}$ energies, calculated at first order by equation (\ref{esofirst}), compared
with the exact numerical result.

 The numerical value of $E^{SO}_{J^{'}}$ is obtained
calulating numerically the energy eigenvalues of the $4f^8$ Hamiltonian,
and subtracting the ground state energy of the $4f^7$ one, where SO interaction is accounted for in
both Hamiltonians. In the numerical calculation  $F_2,F_4,F_6$ are
taken from Thole\cite{thole}. The parameter $F_0$ is already contained in $\Delta E$, so we take $F_0$ equal to zero 
in numerical calculations.

In figure (1) we show the energies as a function of $\Delta E$ for the antiferromagnetic eigenstates of
$H^{SO}(S_z)$ ( equation (\ref{Heffective_simpleS}) ) for $S^{'}_z$ between $1$ (smallest energy )  and $3$
(highest energy), with a dashed line, compared to the exact numerical
solution ( solid line). The ground state energy for $S^{'}_z=0$ has
been subtracted (it is taken as origin of the energy scale).
In the left panel the $4f^8$ energies entering equation(\ref{Heffective_simpleS}) are calculated at first 
first order in SO, while in the right panel numerical  $4f^8$ energies for an isolated ion are used.
The parameters used in the calculation are $t=0.05eV$ and $\Delta E$ between $0.4$  and $1$.
One can observe that the simple formula  \ref{Heffective_simpleS} is an excellent approximation
when exact energies are considered in the denominator.

The energy splitting has to be compared with the magnetic field strength.
Considering a typical XAS-XMCD  experimental case \cite{denadai} with a $7$ Tesla field,
the energy gain got aligning about $7$ Bohr magnetons from perpendicular to parallel direction  with the field 
is 0.4meV. This energy is of the same order of magnitude of, or lower, than the splitting caused by  hybridisation plus spin-orbit.

This effect can therefore, depending on $t$ and $\Delta E$, prevail on the magnetic polarizing field 
and suppress partially, or completely, the magnetization. 


\begin{table}
\begin{ruledtabular}
\begin{tabular*}{\hsize}{c|ccccccc}
$ $ & $j=6$ & $j=5$ & ... & .. &..  & .. & $j=0$ \\
 eq. \ref{esofirst}     & -0.295  &  -0.098     & 0.065    &   0.197 & 0.295  &0.361 & 0.394\\
$ E_{4f^8}-E_{4f^7}$   & -0.302   &   -0.066   &  0.076   &  0.189  & 0.268   & 0.321 & 0.346  \\
\end{tabular*}
\end{ruledtabular}
\caption{\label{tabSOJcalc}  
  Dependancy of $4f^8$ ground states as a function of total moment J. First order formula ( first line)
is compared to numerical results ( second line ) obtained using
  parameters from reference \cite{thole}. Units are eV.
}
\end{table}
\begin{figure} 
\centering
\includegraphics[angle=-90, totalheight=0.32\textheight,viewport= 0  40 600 800, clip  ]{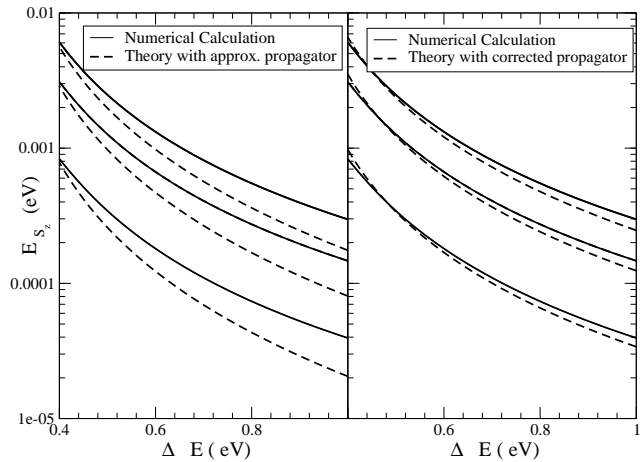}
\caption{
Energies as a function of $\Delta E$ for $S^{'}_z$ between $1$ (smallest energy )  and $3$
(highest energy).   The ground state  for $S^{'}_z=0$ is taken as origin of the energy scale. The dashed line is equation  (\ref{Heffective_simpleS}) 
with  $4f^8$ energies energies at first order in SO (left) and calculated numerically
for an isolated ion (right). Comparison is made with  the exact numerical solution ( solid line).
}
\label{fig1}
\end{figure}

\section{Discussion and Conclusions}

We have shown in the previous section that a very small anisothropic
hybridisation ( $t=0.05$ eV ) can give, for weak magnetic fields, a complete suppression of the
magnetisation  along  the encaged metal displacement axis. At zero
temperature the magnetization would be a discontinuous function of the
magnetizing field. For a polarizing field perpendicular to the
displacement axis the magnetization curve would be instead continuous.

In a real system one should take into account temperature and
disorder. Temperature effect would tend to smear discontinuities.

Disorder may be of different kinds. One  kind of disorder  are
 fluctuations in $\Delta E$ due to inhomogeneities of cage
environments. As $\Delta E$  affects the
propagator denominator of equation (\ref{firstordersimple})
fluctuations  might influence greatly the experimental
result: discontinuities could be smeared out because  the moment of cages having lower
$\Delta E$ is depressed more than that of higher $\Delta E$ ones.

Disorder of the displacement axis direction in the sample would have a
similar effect.

These consideration could explain why experiments show  magnetization
curves that are continuous and  saturate at reduced values.

\begin{figure} 
\centering
\includegraphics[angle=-90, totalheight=0.32\textheight,viewport= 0  40 600 800, clip  ]{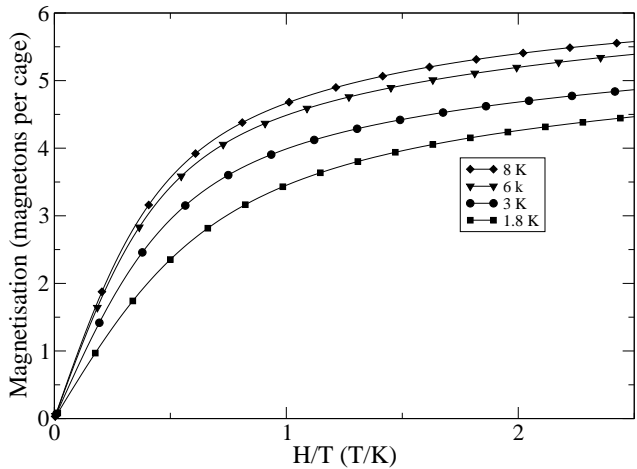}
\caption{
Magnetisation   $1.8$,$3$,$6$ and $8$K temperature
for  the parameters choice $\Delta E=1eV$ and $t=0.2eV$.
}
\label{fig2}
\end{figure}

As an example we calculate magnetisation curves at different temperature
in the case of random orientation of the displacement axis.
We consider the parameters $\Delta E=1eV$ and $t=0.2eV$.
The magnetisation is shown in figure 2 for $1.8$,$3$,$6$ and $8$K.
These data have to be compared with figure 8 of reference\cite{hua1}.
The experimental behaviour is reproduced.

The above discussion leaves the problem still open. First 
of all  further investigation is needed to better evaluate
the real values of $\Delta E$ and $t$,  that in this work
we have chosen arbitrarily with the only criteria of  giving  a numerical example
based on  conservative values ( small $t$ and non negligeable $\Delta E$).

Second, a comparison with a complete set of data using 
a realistic modelisation of the sample should be done.

However we can conclude that the effect is important 
even for very small hybridization, and therefore
cannot be ignored if one wants really to understand the magnetic
properties of encaged RE.

\begin{acknowledgments}

I am grateful to the people of the ID08 beamline at ESRF, in
particular Nick Brookes and Celine De Nadai for introducing me to this
subject and  motivating this analysis, and for the very fruitful discussions. 

\end{acknowledgments}

\end{document}